\documentclass[nofootinbib,twocolumn,showpacs,showkeys]{revtex4}
\usepackage{graphicx,color}
\usepackage{amssymb}
\usepackage{amsmath}
\usepackage{bm}
\usepackage{lipsum}
\usepackage{latexsym}
\usepackage{dcolumn}
\usepackage{float}
\usepackage{hyperref}

\begin{document}

\title{Equilibrium and stability of relativistic stars in extended theories of gravity}

\author{Aneta Wojnar$^{1,2,3}$}\email{aneta.wojnar@ift.uni.wroc.pl}
\author{Hermano Velten$^{4}$}\email{velten@pq.cnpq.br}

\affiliation{$^1$Institute of Physics, Maria Curie-Skłodowska University, Plac Marii Curie-Skłodowskiej 1, 20-031 Lublin, Poland}
\affiliation{$^2$INFN Sez. di Napoli, Univ. di Monte S. Angelo, Ed. G, Via Cinthia, I-80126 Napoli, Italy.}
\affiliation{$^3$Dipartimento di Fisica "E. Pancini", Universita' di Napoli Federico II, Complesso
Universitario di Monte S. Angelo, Via Cinthia, 9, I-80126 Naples, Italy.}
\affiliation{$^4$Universidade Federal do Esp\'{\i}rito Santo (UFES), Av. Fernando Ferrari S/N, 29075-910, Vit\'oria, Brazil}

\begin{abstract}
We study static, spherically symmetric equilibrium configurations in extended theories of gravity (ETG) following the notation introduced 
by Capozziello et {\it al}. We calculate the differential equations for the stellar structure in such theories in a very generic form i.e., the 
Tolman-Oppenheimer-Volkoff generalization for any ETG is introduced. Stability analysis is also investigated with special focus on the particular example
of scalar-tensor gravity.

\end{abstract}

\maketitle

\section{Introduction}

Classical General Relativity (GR) is a very elegant theory which is, roughly speaking, 
described by the Einstein's field equations. They show the relation between geometry of spacetime and matter (fields) contribution. 
Many astronomical observations have tested and confirmed GR predictions both at solar system level and using binary pulsars as well as the recent positive
detection of gravitational waves. Therefore, one deals with convincing indication that the GR is the theory responsible for mediating the gravitation interaction. However, 
it seems that building a successful model for the dynamics of the universe using GR and known matter fields as the source of Einstein equations is not enough 
to describe many issues that recently appeared in fundamental physics, astrophysics and cosmology. There are compelling evidences from recent observations 
leading to the so called dark energy phenomena \cite{sami, hut}, i.e., a late-time cosmic acceleration (which one explains by adding an exotic fluid called 
dark energy) as well as the dark matter puzzle \cite{report, cap_far}. Also, an important ingredient of the concordance model is an inflationary phase in the
early universe \cite{starobinsky, guth} supposedly driven by an inflaton field. The common point here is that unknown components should be added to the
energy momentum tensor when GR is adopted. While the direct confirmation about the existence of such extra fields is missing, one alternative relies in 
assuming that GR is not the actual gravitational theory acting on large scales but is only recovered locally via screening mechanisms \cite{khoury} (although there is evidence that screening solutions can eventually fail inside screened regions \cite{Piazza} and astrophysical objects\cite{screen2}). Then, the observations leading to the dark energy/matter phenomena would result from some non-trivial prediction of new gravitational theories. Hence, the search for new approaches for the gravitational interaction has become a fruitful investigation route. 

%Also, an additional argument in a favor of searching for
%generalizations of gravity is the non-renormalizibility of Einstein's theory \cite{stelle}. One would also like
%to  unify  gravity  theory  with  the  other  ones   (electromagnetism,  weak  and  strong interactions  has
%been already unified into the Standard Model) but no satisfactory result has been obtained so far
%(string theory, supersymmetry) that could combine particle physics and gravitation.

There are also problematic issues concerning astrophysical objects like for instance neutron stars. Their structure and the relation between the mass and 
the radius are determined by equations of state (EoS) of dense matter. The maximal mass value of such objects is still an open question but recent observations 
estimate this limit as $2$M$_\odot$. For example, the pulsar in the system PSR J0348+0432 has the limit $2.01$M$_\odot$ \cite{Anto}, another massive neutron star are Vela X-1 with the mass $\sim 1.8$M$_\odot$ \cite{rawls} and (B1957+20) with a mass $\sim 2.4$M$_\odot$ \cite{van}. A compilation with recent neutron stars mass determination can be found in \cite{stellarcollapse}. It should be also mentioned that some EoS include hyperons which make the maximal mass limit for non-magnetic neutron stars 
significantly lower than 2M$_\odot$ \cite{hyperon}. Therefore, very massive neutron stars challenge the viability of hyperon based EOS. 
There are a few ways to approach the ``hyperon puzzle", such as hyperon-vector coupling, chiral quark-meson coupling and the existence of strong magnetic 
fields inside the star. Concerning the latter approach, for example, some works find a clear influence of the magnetic field increasing the mass of the star \cite{mag}. On the other hand, it seems that the existence of neutron stars---without strong magnetic fields---with masses larger than $2$ M$_{\odot}$ is challenged within the framework of GR \cite{astash, astash2, astash3}. Also, it is worth noting that the usual formalism about the
effects of density-dependent magnetic fields on the properties of neutron stars has been shown to be controversial \cite{controversialmag}.

As neutron stars are very peculiar objects for testing 
theories of matter at high density regimes, data about their macroscopic properties like mass and radius can also be used for studying potential deviations from GR. There exist suggestions \cite{astash3, eksi} that the use GR, if adopted to describe strong gravitational fields, is only a phenomenological extrapolation since the strength of gravity sourced by a neutron star is many orders of magnitude larger that the one probed by the solar system (weak field limit) tests. Also following this line of reasoning, from theoretical and experimental reasons one believes that GR 
should be modified when gravitational fields are strong and spacetime curvature is large \cite{berti}. Therefore, a promising route of investigation is firstly to set a specific model of dense matter, i.e., an equation of
state, and then proceed computing macroscopic properties of neutron stars in a given ETG. Indeed, the predictions of alternative theories to GR concerning the structure of compact objects is currently an active research field \cite{AltNS}.   

In order to implement the strategy described at the end of the last paragraph the first step is to obtain the equilibrium configuration for the stellar objects. In GR, the simplest case of a static, spherically symmetric geometry gives rise to the well known Tolmann-Oppenheimer-Volkoff (TOV) equation \cite{OV, Tolman:1939dn}. Our goal in this work is twofold: i) firstly, we introduce a TOV-like equation for a general class of ETG (as studied in \cite{mim, mim2, mim3}) which will be presented in section II. The generalized equilibrium equation for any ETG (the TOV-like equation) is then calculated; ii) We also generalize the stability criterion for ETG in chapter III. A parametrized version of the TOV equation has been presented in Ref. \cite{fp}. We extend the analysis of the stability based on the thermodynamical properties of the system as presented in some textbooks. See for example\cite{weinberg}. As a case study we investigate the stability conditions for scalar-tensor gravity. We work in $c=1$ units with the signature $(-+++)$ for the metric.

\section{Extended theories of gravity and stellar equilibrium configurations}
We denote by extended theories of gravity (ETG) any alternative to GR in which the field equations can be recast in the form \cite{mim, mim2, mim3}
\begin{equation}\label{mod1}
 \sigma(\Psi^i)(G_{\mu\nu}-W_{\mu\nu})=\kappa T_{\mu\nu},
\end{equation}
where $G_{\mu\nu}=R_{\mu\nu}-\frac{1}{2}Rg_{\mu\nu}$ is the Einstein tensor, $\kappa=-8\pi G$, the factor $\sigma(\Psi^i)$ is a coupling to the gravity
while $\Psi^i$ represents for instance curvature invariants or other fields, like 
scalar ones. The symmetric tensor $W_{\mu\nu}$ stands for additional geometrical terms which may appear in an specific ETG under consideration. It is important
to note that (\ref{mod1}) represents a parameterization of gravitational theories at the level of field equations. 
The energy-momentum tensor $T_{\mu\nu}$ will be considered as the one of a perfect fluid, that is $T_{\mu\nu}=pg_{\mu\nu}+(p+\rho)u_\mu u_\nu$, where $p$ and 
$\rho$ are the pressure and the energy density of the fluid. The four velocity $u^\mu$ of the co-moving (with the fluid) observer is normalized 
with the condition $u^\mu u_\mu = -1$. 

It is worth noting that (\ref{mod1}) does not encompass all the possible alternatives to GR at the field equations level. However, most of 
the main proposals like, for instance, scalar tensor theories and $f(R)$, can be reshaped in this
form. As an example, for theories which have a time dependent effective gravitational coupling $\sigma \equiv \sigma(t)$ and $W_{\mu\nu}=0$. 

One may also add a coupling to the matter source (as it appears often in the so-called Einstein frame) but here we will not consider that case. From the
structure of (\ref{mod1}) one sees that GR is immediately recovered if $\sigma(\Psi^i)=1$ and $W_{\mu\nu}=0$. The extended Einstein's field 
equations (\ref{mod1}) can also be written as
\begin{equation}\label{mod}
G_{\mu\nu}=\kappa T^{eff}_{\mu\nu}=\frac{\kappa}{ \sigma} T_{\mu\nu}+W_{\mu\nu}.
\end{equation}
It is worth noting that one cannot postulate that the energy-momentum tensor of the matter $T_{\mu\nu}$ is conserved. Rather, due to 
the Bianchi identity, the effective energy-momentum tensor $T^{eff}_{\mu\nu}$ is conserved i.e., $\nabla_\mu T_{eff}^{\mu\nu}=0$. In some special ETG cases \cite{koivisto} one deals with modifications of the conservation of the matter energy-momentum tensor.

\subsection{TOV equations in ETGs}

The simplest configuration for a star is the static and spherically symmetric geometry as given by the metric
\begin{equation}
 ds^2=-B(r)dt^2+A(r)dr^2+r^2d\theta^2+r^2\sin^2{\theta} d\phi^2.
\end{equation}

From the normalization condition one has that $u_0=-\sqrt{B(r)}$. As the metric is time independent and spherically symmetric, the pressure $p$ and energy 
density $\rho$ are functions of the radial coordinate $r$ only. Hence we will assume that the coupling function $\sigma$ and 
the geometric contributions $W_{\mu\nu}$ are also independent of the coordinates $(t,\theta,\phi)$. 

We calculate in detail the components of (\ref{mod}). The components of the Ricci tensor read
\begin{align}
 R_{tt}&=-\frac{B''}{2A}+\frac{B'}{4A}\left(\frac{A'}{B}+\frac{B'}{B} \right)-\frac{B'}{rA}\\
&=\frac{\kappa}{2\sigma}(\rho+3p)B+W_{tt}+\frac{BW}{2},\\
 R_{rr}&=\frac{B''}{2A}-\frac{B'}{4B}\left(\frac{A'}{B}+\frac{B'}{B} \right)-\frac{A'}{rA}\\
&=\frac{\kappa}{2\sigma}(p-\rho)A+W_{rr}-\frac{AW}{2},\\
 R_{\theta\theta}&=-1+\frac{r}{2A}\left(-\frac{A'}{B}+\frac{B'}{B}\right)+\frac{1}{A}\\
&=\frac{\kappa}{2\sigma}(p-\rho)r^2+W_{\theta\theta}-\frac{r^2W}{2},
\end{align}
where $W=-B^{-1}W_{tt}+A^{-1}W_{rr}+2r^{-2}W_{\theta\theta}$ is the trace of the tensor $W_{\mu\nu}$. The symbol prime $(^{\prime})$ denotes the 
derivative with respect to $r$.
Using the above equations to write
\begin{equation}
 \frac{R_{rr}}{2A}+\frac{R_{00}}{2B}+\frac{R_{\theta\theta}}{r^2}=-\frac{A'}{rA^2}-\frac{1}{r^2}+\frac{1}{Ar^2}=\frac{\kappa\rho}{\sigma}+r^2B^{-1}W_{tt}
\end{equation}
we obtain the following relation
\begin{equation}\label{forA}
 \left( \frac{r}{A} \right)'=1+\kappa  r^2\frac{\rho(r)}{\sigma(r)}+r^2B^{-1}(r)W_{tt}(r).
\end{equation}
Then we may solve equation (\ref{forA}) and write the solution in the very familiar form
\begin{equation}\label{mod_geo}
 A(r)=\left( 1-\frac{2G \mathcal{M}(r)}{r} \right)^{-1},
\end{equation}
where the mass function $\mathcal{M}(r)$ is defined here as
\begin{equation}
\mathcal{M}(r)=\int^r_0\left( 4\pi \tilde{r}^2\frac{\rho(\tilde{r})}{\sigma(\tilde{r})}-\frac{\tilde{r}^2W_{tt}(\tilde{r})}{2GB(\tilde{r})} \right)d\tilde{r}.
\end{equation}
This solution is clearly different from the usual definition given by GR in which $\mathcal{M}(R)$ ($R$ is the radius of the star) is interpreted as the physical mass of the central object. Here, this expression should be interpreted as the mass function of the coupled TOV-like system. 
Geometric quantities also enters here. This expression is different from the actual physical mass, the one inferred from binary pulsar
observations, for instance, and used to plot the usual {\it Mass-Radius} diagram.

The complete derivation also needs the relations
\begin{align}
 \frac{A'}{A}=\frac{1-A}{r}-\frac{\kappa Ar}{\sigma}Q,\label{for1}\\
 \frac{B'}{B}=\frac{A-1}{r}-\frac{\kappa Ar}{\sigma}\Pi,\label{for2}
\end{align}
where we have defined  new quantities
\begin{eqnarray}\label{def}
Q(r):=\rho(r)+\frac{\sigma(r)W_{tt}(r)}{\kappa B(r)},\\
\label{def2} \Pi(r):=p(r)+\frac{\sigma(r)W_{rr}(r)}{\kappa A(r)}.
\end{eqnarray}
In GR the functions $Q$ and $\Pi$ would be interpreted as the energy density and pressure, respectively.

The conservation of the effective energy-momentum tensor is another useful relation. The hydrostatic equilibrium $\nabla_\mu T^{\mu\nu}_{eff}=0$ reads then
\begin{equation}\label{equil}
\kappa(\sigma^{-1}\nabla_\mu T^{\mu\nu}-\sigma^{-2}T^{\mu\nu}\nabla_\mu\sigma)+\nabla_\mu W^{\mu\nu}=0,
\end{equation}
or, more explicitly,
\begin{align}
&\kappa\sigma^{-1}\left(p'+(p+\rho)\frac{B'}{2B}\right)-\kappa p\frac{\sigma'}{\sigma^2}-\frac{A'}{A^2}W_{rr}+A^{-1}W'_{rr}\nonumber \\
&+\frac{2W_{rr}}{Ar}+
\frac{B'}{2B}\left(\frac{W_{rr}}{A}+\frac{W_{tt}}{B}\right)-\frac{2W_{\theta\theta}}{r^2}=0.
\end{align}
Let us notice that from (\ref{def2}) and with the help of (\ref{for1})
\begin{equation}\label{tov1}
 \left(\frac{\Pi}{\sigma}\right)'=\frac{p'}{\sigma}-\frac{p\sigma'}{\sigma^2}+\frac{W'_{rr}}{\kappa A}-\frac{W_{rr}(1-A)}{\kappa rA}+\frac{rW_{rr}Q}{\sigma}.
\end{equation}
This equation is the basic structure for deriving the generalized hydrostatic equilibrium for stars in ETG. Together with (\ref{for2}) and definition (\ref{def}), 
the equation (\ref{tov1}) can be written as
\begin{eqnarray}\label{tov}
  \left(\frac{\Pi}{\sigma}\right)'&=&-\frac{Gm}{r^2}\left(\frac{Q}{\sigma}+\frac{\Pi}{\sigma}\right)
  \left(1+\frac{4\pi r^3\frac{\Pi}{\sigma}}{\mathcal{M}}\right)\left(1-\frac{2G\mathcal{M}}{r}\right)^{-1}\nonumber\\ 
	&+&\frac{2\sigma}{\kappa r}\left(\frac{W_{\theta\theta}}{r^2}-\frac{W_{rr}}{A}\right).
\end{eqnarray}

The above equation (\ref{tov}) and 
\begin{equation}\label{mr}
\mathcal{M}(r)= \int^r_0 4\pi \tilde{r}^2\frac{Q(\tilde{r})}{\sigma(\tilde{r})} d\tilde{r}.
\end{equation}
have a similar functional form as the standard GR result. But one remarkable difference is the existence of the geometrical contribution in the last term of (\ref{tov}). The set of equations (\ref{tov}) and (\ref{mr}) represent a useful tool for studying stellar configurations once a certain ETG is specified.

It is worth noting that such equations determine completely the stellar equilibrium since the 
assumption that the pressure is expressed as a function of the density only, i.e., the entropy per nucleon and the chemical composition as constant throughout 
the star. Such assumptions will also be used in the analysis of stability of these systems.

It should be noticed that since the tensor $W_{\mu\nu}$ can include some extra fields like scalar ones for example, 
besides the generalized Einstein's field equations (\ref{mod1}), one will inevitably deal with equations of motion for the additional fields which should be taken into account. This means that Eq. (\ref{tov}) and (\ref{mr}) are general up to the definition of the specific theory. After that stage, these equations can be further simplified with the help of the new equations of motion of the specified theory. In order to exemplify such issue, in the next section we will consider scalar-tensor gravity where the modified Klein-Gordon equation for the scalar field is taken into account.

\section{Stability conditions}

In order to obtain specific predictions on the stability of static and spherically symmetric systems within the general form like (\ref{mod1}) one has to provide the $W_{\mu \nu}$ and $\sigma(\Psi^i)$ terms. In this section, as a case study, we calculate the stability criterion for the $k$-essence class of theories.

\subsection{Extended stability conditions applied to scalar-tensor gravity}

In scalar-tensor theories the gravitational interaction is mediated not only by the metric field (as in GR), but also for a scalar field $\phi$. Among many realizations of scalar-tensor theories, a simple prototype is the quintessence class in which the scalar field is said to be minimally coupled to the geometrical sector. 

The theory can be written according to the following action
\begin{equation}
 S=\frac{1}{2\kappa}\int d^4x\sqrt{-g}( R-\nabla_\mu\phi\nabla^\mu\phi-2V(\phi))+S_m[g_{\mu\nu},\psi].
\end{equation}

The field equations derived from the above action are
\begin{eqnarray}\label{fieldEQ}
 G_{\mu\nu}+\frac{1}{2}g_{\mu\nu}\nabla_\alpha\phi\nabla^\alpha\phi-\nabla_\mu\phi\nabla_\nu\phi+g_{\mu\nu}V(\phi)&=&\kappa T_{\mu\nu},\label{st1}\\
 V'(\phi)-\Box\phi&=&0.
\end{eqnarray}
Since we are working in a curved spacetime the scalar field $\phi$ depends on matter contribution ($\rho$) via the d'Alembertian ($\Box$) operator present in the modified Klein-Gordon equation above.

Comparing (\ref{fieldEQ}) with (\ref{mod1}) one notice that in the $k$-essence case we identify $\sigma=1$ and  
\begin{equation}
W_{\mu\nu}=-\frac{1}{2}g_{\mu\nu}\nabla_\alpha\phi\nabla^\alpha\phi+\nabla_\mu\phi\nabla_\nu\phi-g_{\mu\nu}V(\phi).
\end{equation}
From the above definition of $W_{\mu\nu}$ we can write the following components
\begin{align}
 W_{tt}&=\frac{1}{2}B \nabla_\alpha\phi\nabla^\alpha\phi +BV(\phi)=B(C+2V),\\
 W_{rr}&=AC,\\
 W_{\theta\theta}&=-r^2 (C+2V).
\end{align}
In the above expressions we have defined $V\equiv V(\phi)$ and 
\begin{equation}\label{C}
C\equiv C(Q, \phi,\phi')=\frac{1}{2}A^{-1}\phi'^2-V. 
\end{equation}
Let us recall that $A$ is a function of $Q$.
Hence, the last term in the generalized TOV equation (\ref{tov}) is $-\frac{4\sigma}{\kappa r}(C+V)=-2\sigma\frac{\phi'^2}{\kappa Ar}$. Moreover, 
in the $k-$essence case, the functions $Q$ and $\Pi$ will assume the form
\begin{align}
\label{Qk} Q_k&=\rho(r)+\kappa^{-1}(C+2V),\\
 \Pi_k&=p(r)+\kappa^{-1}C.
\end{align}
%\textcolor{red}{see Weinberg, page 49. This is done for special relativity but 
 %he uses that result for GR, that is, NS's.} 
Let us calculate in detail the stability analysis. We assume that the particle number $N^\alpha=n u^\alpha$ is conserved
 \begin{equation}
  \nabla_\alpha(n u^\alpha)=u^\alpha \nabla_\alpha n + n\nabla_\alpha u^\alpha=0.
 \end{equation}
The crucial issue here is that we are dealing with the effective energy-momentum tensor (from the Bianchi identities $\nabla_\mu G^{\mu\nu}=0$), therefore
\begin{eqnarray}\label{thermo}
u_\nu \nabla_\mu T^{\mu\nu}_{eff} &=&\sigma^{-1} \left(u^\mu\nabla_\mu p-n u^\mu\nabla_{\mu}\left(\frac{p+\rho}{n}\right)+\rho u^\mu\nabla_\mu\sigma\right) \nonumber\\
&+&u_\nu\nabla_\mu W^{\mu\nu},
\end{eqnarray}
and
\begin{eqnarray}\label{eq_therm}
u^\mu\left(\frac{\sigma}{\kappa} W^\nu_{\;\mu;\nu} -n p\nabla_\mu\left(\frac{1}{n}\right)-n\nabla_\mu\left(\frac{\rho}{n}\right)+\frac{\rho}{\sigma}\nabla_\mu\sigma \right)=0.
\end{eqnarray}

%\begin{eqnarray}\label{eq_therm}
%-n u^\mu\left( p\nabla_\mu\left(\frac{1}{n}\right)+\nabla_\mu\left(\frac{\rho}{n}\right)+\frac{\rho}{n}\nabla_\mu\sigma \right) +\sigma u^\mu W^\nu_{\;\mu;\nu}=0.
%\end{eqnarray}
Since we are working with the modified field equations (\ref{st1}), the coupling term $\nabla_\mu\sigma$ in the above formula vanishes.
Keeping in mind that the tensor $W_{\mu\nu}$ does not depend
on the energy density, the only non-vanishing terms that undergo infinitesimal changes with respect to the infinitesimal changes of the energy density are 
\begin{equation}
\delta\left( \frac{\rho}{n} \right)+p\delta\left(\frac{1}{n} \right)=0,
\end{equation}
and consequently,
\begin{equation}\label{deltanr}
 \delta n(r) =\frac{n(r)}{p(r)+\rho(r)}\delta\rho(r).
\end{equation}
Notice that the second law of thermodynamics will differ in ETG's \cite{bamba} but the above general relation remains valid in our treatment. This expression will be our starting point but before doing that we should investigate in detail the dependence of the variation $\delta \rho$ with respect to other quantities. 
It is important to notice the particular form of the effective energy-momentum tensor (\ref{fieldEQ}) in the ETG that we are considering here. As $W_{\mu\nu}$ is symmetric and one also deals with the modified K-G 
equation then
\begin{equation}
 \nabla^\mu W_{\mu\nu}=\nabla^\mu\phi(\nabla_\mu\nabla_\nu\phi-\nabla_\nu\nabla_\mu\phi)=0
\end{equation}
where we have used the K-G equation $\Box\phi=V'$. One may also compute it explicitly for the component $\mu=r$
\begin{equation}
 \nabla_\nu W^\nu_{\;r}=C'+(C+V)(\frac{A-1}{r}-\kappa Ar\Pi+\frac{4}{r}):=C'+D.
\end{equation} 

Applying the Klein-Gordon equation to the derivative $C'=\frac{dC(\phi,\phi')}{dr}$, gives rise to the expression $C'=-(C+V)(\frac{A-1}{r}-\kappa Ar\Pi+\frac{4}{r})=-D$ which will be useful latter. Therefore, component $\mu=r$ of the equation (\ref{eq_therm}) resembles the GR form
\begin{equation}\label{new}
 n'(r)=n\frac{\rho'}{\rho+p}.
\end{equation}
Now on, we are going to use the Lagrange multipliers method following the procedure presented, for example, in Ref. \cite{weinberg}. The nucleon number 
\begin{equation}
N=\int^R_0 4\pi r^2 [1-2G\mathcal{M}(r)/r]^{-1/2}n(r)dr
\end{equation}  
remains unchanged but we should remember that it also depends on the modified geometry (see the formula (\ref{mod_geo}) and below). Then, we find
\begin{eqnarray}\label{deltaMlambda}
 &0&=\delta \mathcal{M}-\lambda \delta N=\int_0^\infty 4\pi r^2 \delta Q dr \nonumber\\
 &-&\lambda \int_0^\infty 4\pi r^2\left(1-\frac{2G\mathcal{M}(r)}{r}\right)^{-\frac{1}{2}}\delta n(r)dr\nonumber\\
 &-&
 \lambda G\int_0^\infty 4\pi r\left(1-\frac{2G\mathcal{M}(r)}{r}\right)^{-\frac{3}{2}}n(r)\delta\mathcal{M}(r)dr,
\end{eqnarray}

 In the above expression one has to identify explicitly the terms $\delta Q_{\kappa}$ and $\delta n(r)$ since they depend on the variations of geometrical quantities as well as $\delta\rho(r)$. Therefore, the variation of each term e.g., $\delta n \equiv \delta n (\delta Q_k, \delta C)$ should be written in details.

From (\ref{Qk}) one realizes that the variation $\delta\rho(r)$ can be expressed as
\begin{equation}
\delta\rho =\delta Q_{k} -\kappa^{-1}(\delta C + 2 V' \delta\phi),
\end{equation}
where $\delta C$ is a function of $\delta Q_k$, $\delta\phi$, and $\delta\nabla^\mu\phi$. Our goal is to show that the equilibrium is stable with respect to radial oscillations if $\mathcal{M}$ is a minimum with respect to all possible variations. 

%Therefore, now from Eq. (\ref{deltaMlambda}) we should examine the variation of $\delta n$ in details.

In astrophysical application, temporal variations of the scalar field can be neglected. Then, since the scalar field is a function of the radial coordinate only, i.e., $\phi\equiv\phi(r)$, we may write $\phi'=\partial_\mu\phi=\nabla_\mu\phi$. From (\ref{C}) the term $\delta C$ is written as
\begin{equation}
 \delta C=-\frac{G}{r}\phi'^2\int^R_0 (4\pi r^2 \delta Q_k dr)+\nabla_\mu\phi\delta\nabla^\mu\phi-\Box\phi\delta\phi,
\end{equation}
where we have used the K-G equation $\Box\phi=V'$. Therefore, the full expression for the variation of the energy density becomes
\begin{equation}
 \delta\rho =\delta Q_{k} -\kappa^{-1}\left(-\frac{G}{r}\phi'^2\int^R_0 (4\pi r^2 \delta Q_{k} dr)+\nabla_\mu(\delta\phi\nabla^\mu\phi)\right).
\end{equation}

This quantity indicates how the thermodynamical relation (\ref{deltanr}) is modified in the presence of the scalar field $\phi$. Moreover,
we identify the quantity $\delta\mathcal{M}(r)$ which appears in (\ref{deltaMlambda}) as $\delta\mathcal{M}(r')=\int_0^\infty 4\pi r'^2 \delta Q_{k} dr'$. Applying such relations
into equation (\ref{deltaMlambda}) it is worth noting that such equation becomes

\begin{widetext}
 \begin{align}\label{Eq46}
 0=&\delta \mathcal{M}-\lambda\delta N=\int_0^\infty 4\pi r^2 \delta Q_k dr
 -\lambda G\int_0^\infty 4\pi rA^{\frac{3}{2}}n(r)\int_0^\infty (4\pi \tilde{r}^2 \delta Q_k d\tilde{r})dr\nonumber\\
 -&\lambda \int_0^\infty 4\pi r^2A^{\frac{1}{2}}
 \frac{n(r)}{p(r)+\rho(r)}\left(\delta Q_k -\kappa^{-1}\left(-\frac{G}{r}\phi'^2\int^R_0 (4\pi \tilde{r}^2 \delta Q_k d\tilde{r})+\nabla_\mu(\delta\phi\nabla^\mu\phi)\right)\right)dr.
\end{align}
\end{widetext}
Our next step is to investigate carefully the terms appearing in the above expression. For example, one realizes that the last term appearing in (\ref{Eq46})  
 \begin{equation}
  \int_0^\infty 4\pi r^2A^{\frac{1}{2}}\frac{n(r)}{p(r)+\rho(r)}\nabla_\mu(\delta\phi\nabla^\mu\phi)dr,
 \end{equation}
is proportional to the integral of the term $\nabla_\mu(\delta\phi\nabla^\mu\phi)$. This term can be integrated by parts. Then, writing it explicitly after the integration one obtains the following terms
\begin{widetext}
\begin{align}\label{Eqstar}
    &\int_0^\infty\nabla^\mu \left( 4\pi r^2A^{\frac{1}{2}}\frac{n(r)}{p(r)+\rho(r)} \delta\phi\nabla_\mu\phi \right)dr
    -\int_0^\infty \delta\phi\nabla^\mu\phi \nabla_\mu\left( 4\pi r^2A^{\frac{1}{2}}\frac{n(r)}{p(r)+\rho(r)} \right)dr \\
    =& \int_0^\infty\partial^\mu \left( 4\pi r^2A^{\frac{1}{2}}\frac{n(r)}{p(r)+\rho(r)} \delta\phi\partial_\mu\phi \right)dr
  + \int_0^\infty\left[4\pi r^2A^{\frac{1}{2}} \Gamma^\mu_{\mu\nu} \frac{n(r)}{p(r)+\rho(r)} 
  -\partial_\nu\left( 4\pi r^2A^{\frac{1}{2}}\frac{n(r)}{p(r)+\rho(r)}\right) \right] \delta\phi\partial^\nu\phi dr. \nonumber
\end{align}
\end{widetext}
It is important to realize that
\begin{equation}\label{BT}
\int_0^\infty\partial^\mu \left( 4\pi r^2A^{\frac{1}{2}}\frac{n(r)}{p(r)+\rho(r)} \delta\phi\partial_\mu\phi \right)dr
\end{equation} 
which is present in Eq. (\ref{Eqstar}) is a boundary term. We say in advance that our conclusion on the stability of stellar systems in such particular $k-$essence theory crucially depends on this term. It vanishes since both upper and lower substitutions vanish. The former as both variations at the (finite) boundary vanish and the latter as $r^2$ factor takes it zero while other factors are supposedly finite. Therefore, we continue neglecting this term in our analysis
although it should be rethought again if one wants to consider another theory of gravity rather than the minimally coupled scalar field model with the modified Klein-Gordon equation (\ref{fieldEQ}). The discussion above concerns only the minimal coupling case
since a non-minimal coupling could also produce some non-trivial contribution to such boundary term term. The work on non-minimal case is in progress.

Interchanging the $r$ and $\tilde{r}$ integrals will allow us to write
 \begin{widetext}
 \begin{align}\label{Eq48}
\delta \mathcal{M}-\lambda\delta N&=\int_0^\infty4\pi r^2\left[1-\frac{\lambda n(r)}{p(r)+\rho(r)}A^{\frac{1}{2}}-\lambda G \int_r^\infty4\pi \tilde{r}n(\tilde{r})A^{\frac{3}{2}}d\tilde{r}
  -\lambda G\kappa^{-1}\int^\infty_r4\pi \tilde{r} A^{\frac{1}{2}}\frac{n}{p+\rho} \phi'^2 d\tilde{r} \right]\delta Q(r)dr\nonumber\\
  &-
 \lambda\kappa^{-1}\int_0^\infty\partial^\nu\phi \left[4\pi r^2A^{\frac{1}{2}} \Gamma^\mu_{\mu\nu} \frac{n(r)}{p(r)+\rho(r)} 
  -\partial_\nu\left( 4\pi r^2A^{\frac{1}{2}}\frac{n(r)}{p(r)+\rho(r)}\right) \right] \delta\phi dr=0.
 \end{align}
\end{widetext}
In order to assure that the above equality holds one has to guarantee that all the integral terms vanish, that is, both terms 
containing variations must vanish independently. The term in the first line of (\ref{Eq48}) which is proportional to $\delta Q$ will vanish if
\begin{align}\label{eq1}
  \frac{1}{\lambda}=&\frac{n(r)}{p(r)+\rho(r)}A^{\frac{1}{2}}+
 G\int_r^\infty 4\pi \tilde{r} n(\tilde{r})A^{\frac{3}{2}}d\tilde{r}\nonumber \\ 
 +&G\kappa^{-1}\int^\infty_r4\pi \tilde{r} A^{\frac{1}{2}}\frac{n(\tilde{r})}{p(\tilde{r})+\rho(\tilde{r})}\phi'^2 d\tilde{r},
\end{align}
while the vanishing of the last term demands
\begin{equation}\label{eq2}
 4\pi r^2A^{\frac{1}{2}} \Gamma^\mu_{\mu r} \frac{n(r)}{p(r)+\rho(r)} 
  =\partial_r\left( 4\pi r^2A^{\frac{1}{2}}\frac{n(r)}{p(r)+\rho(r)}\right).
\end{equation}

Let us now concentrate on (\ref{eq1}). Deriving it
with respect to the radial coordinate $r$ (where $\lambda$ is constant) and using the fact that in our case $ n'(r)=n\frac{\rho'}{\rho+p}$ still holds, we find
\begin{align}
-4\pi GrA-\frac{p'}{(p+\rho)^2}+\frac{GA}{p+\rho}(4\pi r Q -&\frac{\mathcal{M}}{r^2})\nonumber \\
-\frac{4\pi r G}{\kappa}\frac{\phi'^2}{p+\rho}=&0.
\end{align}
Applying the following relations
\begin{align*}
 \frac{A-1}{r}=& \,A\frac{2GM}{r^2},\\
 p+\rho=& \,\Pi_k+Q_k - 2\kappa^{-1}(C+V),\\
 A^{-1}\phi'^2 =&\, 2(C+V)
\end{align*}
together with
 \begin{align}\label{Capital}
  \Pi'_k&=p'+\kappa^{-1}C'\nonumber \\
  &=p'-\kappa^{-1}(C+V)\left(\frac{A-1}{r}-\kappa Ar\Pi+\frac{4}{r}\right)
 \end{align}
we are finally able to write 
\begin{align}\label{stabilityTOVkessence}
 \Pi'=-\frac{AGM}{r^2}(\Pi+Q)(1+4\pi r\frac{\Pi}{M})-4\frac{C+V}{\kappa r},
\end{align}
which is exactly the TOV equation (\ref{tov}) for the $k$-essence model. 

We turn now our attention to equation (\ref{eq2}). Writing the gamma term explicitly $\Gamma^\mu_{\mu r}=\frac{2}{r}-\frac{1}{2}\left(\kappa Ar(\Pi+Q)\right)$ and using
 $ n'(r)=n\frac{\rho'}{\rho+p}$ and Eq. (\ref{Capital}), we obtain again exactly (\ref{stabilityTOVkessence}), i.e., the TOV form.

Therefore, the system is stable since the boundary term (\ref{BT}) vanishes. If this would not be the case then we were not able to write down the stability equation in the TOV-like form as presented in section II.

\section{Conclusions}
In this work we have studied extended theories of gravity (ETG) based on the phenomenological field equations (\ref{mod1}). Indeed, this class of theories are not derived via standard variation principle from any know action though this represent a general way to implement modifications of GR at the field equations level.

Our focus was to obtain the stellar equilibrium equations for static, spherically symmetric geometries. In the general relativistic case this set of equations is known as the TOV equations. The main result of this work consists in the system of equations (\ref{tov1}) and (\ref{mr}) which is the analogous version of the TOV equations for any ETG. Such equations can now be further applied to specific gravitational theories. For the particular case shown in (\ref{tov}) the TOV structure is preserved only if one finds a suitable theory in which $W_{\theta \theta}= W_{rr} r^2 / A$ and regarded that we identify $Q$ and $\Pi$, as the effective density and effective pressure, respectively.

Concerning the stability of such systems, we argue that this analysis should be implemented case by case only, i.e., it is difficult to achieve general results without specifying the functions $W_{\mu\nu}$ and $\sigma(\Psi^i)$. As an example showing the applicability of our results, we worked on the specific class of $k-$essence theories in section III. For this case, we generalized the stability theorem found for instance in \cite{weinberg} taking into account the new functions $Q$ ans $\Pi$. We found that the specific minimally coupled scalar-tensor case leads to stable configurations since the boundary term (\ref{BT}) does vanish. The considered example shows that there are other theories of gravitation besides GR in which a neutron star is a stable system. However, it is worth noting that other theories can lead to non-vanishing boundary terms.

If one finds a theory in which the equilibrium (\ref{tov}) is not recovered from the Lagrange multiplier method employed in section III, it is clear that stability criterion should be reinterpreted. 
Contrary to the standard case, even assuming uniform entropy per nucleon and chemical composition, the interpretation of the mass function, and consequently the proper definition of the quantity $\mathcal{M}$, should be identified with effective energy density $Q$. The same interpretation should also be extended to the quantity $\mathcal{M}$ which appears in the definition of the nucleon number $N$.

The investigation of the stability of stellar systems in ETG and other modifications of gravity that cannot be recast in the form (\ref{mod1}) should be further investigated. We will present such analysis in a future work.

\noindent
{\bf Acknowledgement:} We thank CNPq (Brazil) and DEC-2013/09/B/ST2/03455 (Poland) for partial financial supports. AW acknowledges support 
from INFN Sez. di Napoli (Iniziative Specifiche TEONGRAV). The authors would like to thank Andrzej Borowiec, David Alvarez Castillo, David Blaschke, Julio Fabris and Oliver Piattella for helpful discussions.

\end{document}